\documentclass[conference]{IEEEtran}
\IEEEoverridecommandlockouts
\usepackage{cite}
\usepackage{amsmath,amssymb,amsfonts}
\usepackage{algorithmic}
\usepackage{graphicx}
\usepackage{textcomp}
\usepackage{xcolor}
\usepackage{siunitx}

\DeclareMathOperator{\rect}{rect}
\DeclareMathOperator{\sgn}{sgn}

\usepackage{acro}

\DeclareAcronym{CS}{
  short        = {CS},
  long         = {Current-steering}
}

\DeclareAcronym{DAC}{
  short        = {DAC},
  long         = {digital-to-analog converter}
}

\DeclareAcronym{SDR}{
  short        = {SDR},
  long         = {signal-to-distortion ratio}
}

\DeclareAcronym{SFDR}{
  short        = {SFDR},
  long         = {spurious-free dynamic range}
}

\def\BibTeX{{\rm B\kern-.05em{\sc i\kern-.025em b}\kern-.08em
    T\kern-.1667em\lower.7ex\hbox{E}\kern-.125emX}}
\begin{document}

\title{
Improved Analysis of Current-Steering DACs Using Equivalent Timing Errors\\
}

\author{
    \IEEEauthorblockN{
        Daniel Beauchamp\IEEEauthorrefmark{1}\IEEEauthorrefmark{2} and
        Keith M. Chugg\IEEEauthorrefmark{2}}
    \IEEEauthorblockA{
        \begin{tabular}{cc}
            \begin{tabular}{@{}c@{}}
                \IEEEauthorrefmark{1}
                    Jariet Technologies, 103 W Torrance Blvd, Redondo Beach, CA 90277 \\
                \IEEEauthorrefmark{2}
                  Ming Hsieh Department of Electrical Engineering, 
University of Southern California, Los Angeles, California 90089 \\
                    \{dbeaucha, chugg\}@usc.edu
            \end{tabular} 
        \end{tabular}
    }
}

\maketitle

\begin{abstract}
\ac{CS} \acp{DAC} generate analog signals by combining weighted current sources. Ideally, the current sources are combined at each switching instant simultaneously. However, this is not true in practice due to timing mismatch, resulting in nonlinear distortion. This work uses the equivalent timing error model, introduced by previous work, to analyze the signal-to-distortion ratio (SDR) resulting from these timing errors. Using a behavioral simulation model we demonstrate that our analysis is significantly more accurate than the previous methods.  We also use our simulation model to investigate the effect of timing mismatch in partially-segmented \ac{CS}-\ac{DAC}s, i.e., those comprised of both equally-weighted and binary-weighted current sources. 
\end{abstract}


\section{Introduction}
Current-steering DACs are considered to be the de facto solution for transmitters in modern high-speed applications\cite{bib:razavi}, including cellular communication, electronic warfare, and automotive radar. The \ac{CS}-\ac{DAC} generates an analog signal from a digital input sequence by combining current sources, as shown in Fig. \ref{fig:full_seg_cs_dac}. We refer to this as a \textit{fully-segmented} \ac{CS}-\ac{DAC}, since each current source is equally weighted. In contrast, \textit{partially-segmented} \ac{CS}-\ac{DAC}s are hybrid architectures that are comprised of both equally-weighted and binary-weighted current sources. Regardless of the architecture, the ideal output is a perfect zero-order-hold (or staircase-like) representation of the digital input sequence. However, this is not true in practice due to nonlinear distortion caused by various errors. 
 
Typically, errors in \ac{CS}-\ac{DAC}s are classified as either static or dynamic \cite{bib:olieman}. Static errors, which are time-invariant and memoryless, are mainly caused by current source mismatch and are treatable by various calibration techniques \cite{bib:stoops}, \cite{bib:trimming}, \cite{bib:DEM}. Dynamic errors, on the other hand, are more difficult to calibrate because they only appear during switching instants and last for a small fraction of the sample period. Hence, this necessitates calibration circuitry with fine resolution in both amplitude and time \cite{bib:chen_paper}, \cite{bib:daigle}. Timing-related mismatch, for example, causes the current cells to fire at different times\cite{bib:timing_errors}; nominally, they all fire simultaneously at each switching instant. 

Dynamic errors, such as timing-related mismatch, limit the high-frequency performance of the \ac{DAC} \cite{bib:chen_paper}, which makes their analysis critical. Previous research on this topic is presented in \cite{bib:timing_errors} and \cite{bib:doris_book}, where it was proposed that timing errors for each current cell can be lumped into an \textit{equivalent timing error} for each switching instant. This is a key contribution that simplifies the analysis considerably. Under this framework, the \ac{DAC} error is comprised of narrow pulses with amplitudes that are proportional to the difference between consecutive input codes. The \ac{SDR} is then derived as a function of the timing error spread, which is assumed to be the same for each current cell.

In this work, we utilize the equivalent timing error introduced in \cite{bib:timing_errors} and provide a more accurate analysis of the resulting model. The key difference in the analyses is that the approach in  \cite{bib:timing_errors}  implicitly assumed that the timing errors are present during the entire sample period. While this assumption leads to an accurate \ac{SDR} for the Nyquist band, it is not as accurate for the \textit{wideband} \ac{SDR}, i.e., where \textit{all} frequency components of the error are considered. In this work we make no such assumption, resulting in a significantly more accurate expression for the wideband \ac{SDR} (as we confirm with behavioral simulations). The limitations of the equivalent timing error model are stated after characterizing the \ac{SDR} over frequency. In addition, we use the behavioral model to explore the \ac{SDR} for partially-segmented architectures.

The rest of the paper is organized as follows. In Section \ref{sec:background}, we provide background information on fully-segmented \ac{CS}-\ac{DAC}s and the problem of timing-related mismatch. In Section \ref{sec:analysis}, we carry out the \ac{SDR} analysis using the equivalent timing error model and compare it to that in \cite{bib:timing_errors}. In Section \ref{sec:simulatons}, we validate our analysis using a behavioral model, state its limitations, and simulate the \ac{SDR} for partially-segmented architectures. Finally, we conclude the paper in Section \ref{sec:conclusion}.  

\section{Background} \label{sec:background}
\subsection{Fully-Segmented Current Steering DAC}
\begin{figure}[!tb]
\centering
\includegraphics[width = 0.65\linewidth]{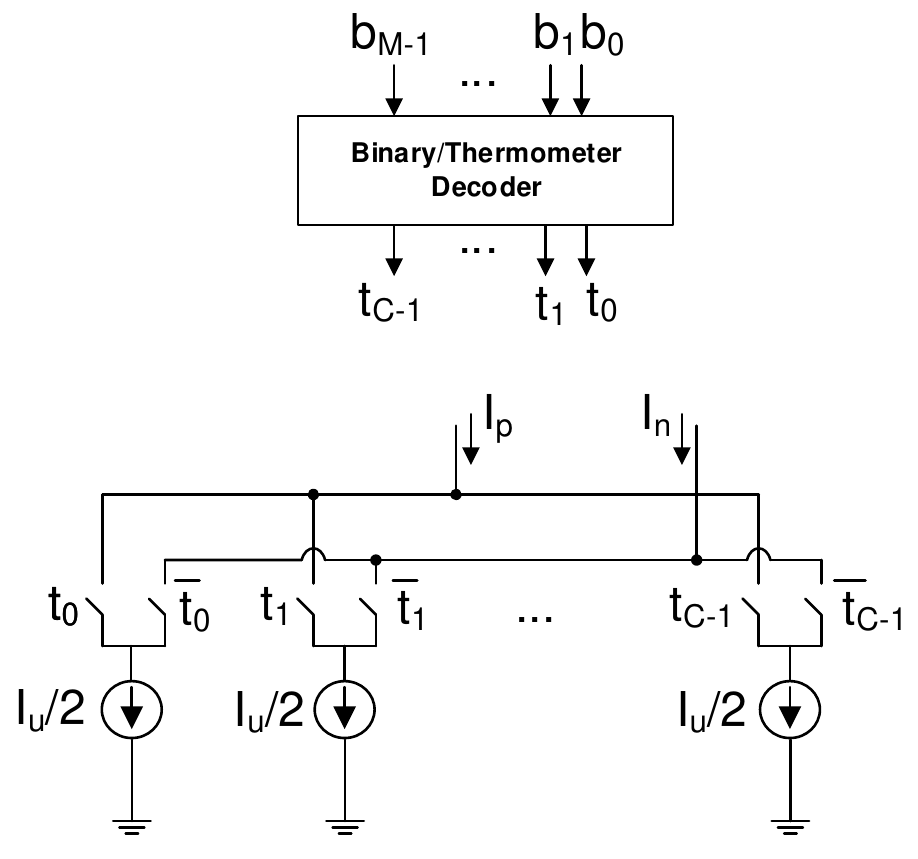}
\caption{Fully-segmented $M$-bit \ac{CS}-\ac{DAC}, $C=2^M-1$.}
\label{fig:full_seg_cs_dac}
\end{figure}

A block diagram of an $M$-bit fully-segmented \ac{CS}-\ac{DAC} is shown in Fig. \ref{fig:full_seg_cs_dac}. 
It is modeled as an array of $C = 2^M-1$ current cells, each weighted by $I_u/2$ with complementary switching. A binary-to-thermometer decoder is used to map the binary input code $b_{M-1} \cdots b_0$ to a thermometer code $t_{C-1} \cdots t_0$. The number of ones in the decoder output is equal to the decimal representation of its binary input, e.g., for a 2-bit \ac{DAC}, $00 \rightarrow 000$, $01 \rightarrow 001$, $10 \rightarrow 011$, and $11 \rightarrow 111$. Each current cell is steered to either the positive or negative output, depending on the input code. In the absence of nonidealities, the \ac{DAC} output is
\begin{align} \label{ideal_dac_output}
    y_\text{ideal}(t) = u(t) * \sum_{n=-\infty}^\infty I_u \, \Delta x_n \, \delta(t-nT_s)
\end{align}
for a digital input sequence $x_n \in \left\{0, \dots, 2^M -1 \right\}$, where $\Delta x_n = x_n - x_{n-1}$, $u(t)$ is the unit step function, and * denotes convolution. 
\subsection{Equivalent Timing Error Model}
In \eqref{ideal_dac_output}, it is assumed that the current cells fire simultaneously at each switching instant, i.e., the \ac{DAC} output changes abruptly at time $nT_s$ when $\Delta x_n \neq 0$. In practice, the $m^\text{th}$ current cell fires at $nT_s + \tau_m$, where $\tau_m$ is the timing error for that cell,  $m \in \left\{0, \dots, C-1 \right\}$. As in \cite{bib:timing_errors}, we model $\tau_m$ as independently drawn from a mean-zero, normal distribution with variance $\sigma^2$ (i.e., $\tau_m \sim N(0,\sigma^2)$). The authors in \cite{bib:timing_errors} begin the analysis by considering the net charge error introduced by the code transition $x_{n-1} \rightarrow x_n$. Next, they formulate an \textit{equivalent timing error} for this transition, $T_\epsilon(n)$, which allows the \ac{DAC} output to be written as
\begin{align} \label{nonideal_dac_output}
    y(t) = u(t) * \sum_{n=-\infty}^\infty I_u \, \Delta x_n \, \delta\left(t-nT_s-T_\epsilon(n)\right)
\end{align}
where
\begin{align} \label{equiv_timing_error}
    T_\epsilon(n) = \begin{cases} \frac{1}{|\Delta x_n|} \, \sum_{m=\min(x_n,x_{n-1})}^{\max(x_n, x_{n-1})-1} \tau_m & \Delta x_n \neq 0 \\ 0 & \Delta x_n = 0
    \end{cases}
\end{align}
\begin{figure}[!tb]
\centering
\includegraphics[width = 0.65\linewidth]{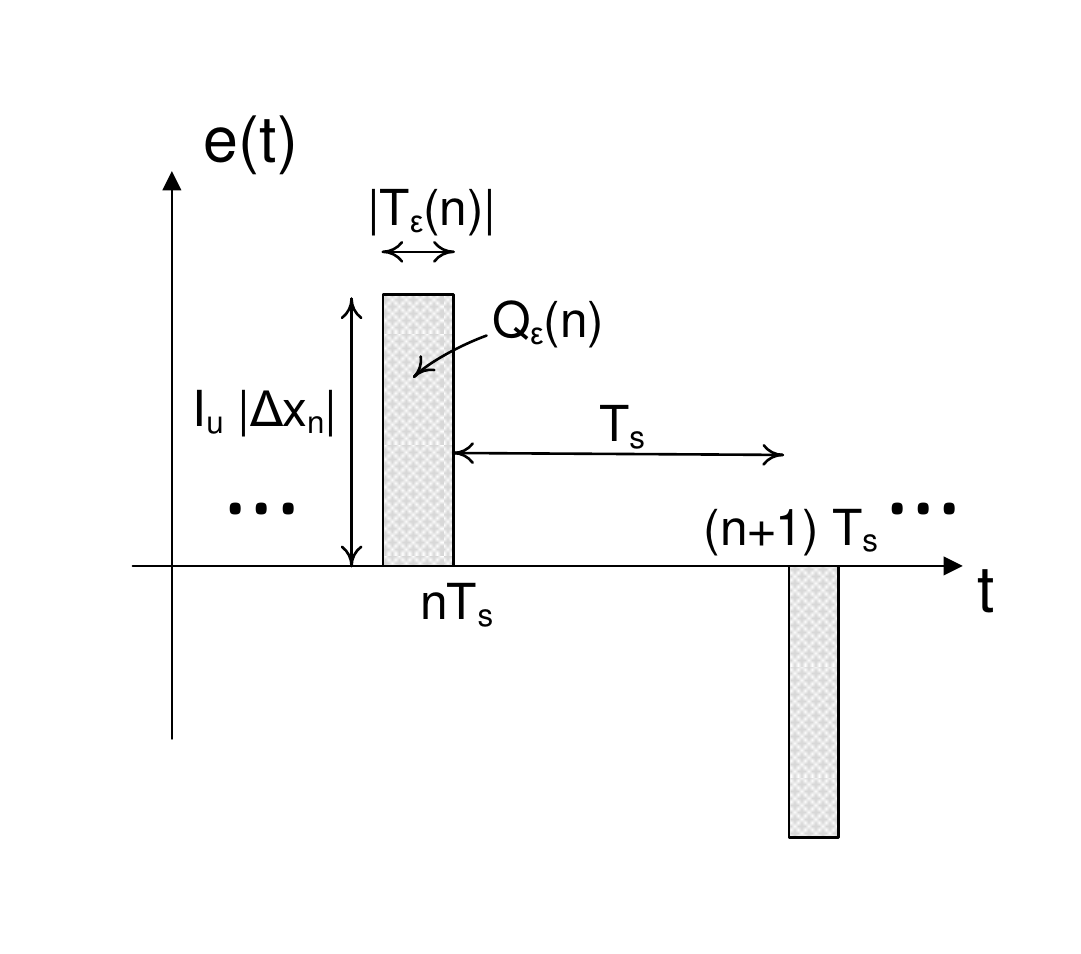}
\caption{Error $e(t)$ based on equivalent timing errors $T_\epsilon(n)$. The charge error $Q_\epsilon(n)$ is the area under the $n^\text{th}$ error pulse.}
\label{fig:error_pulses}
\end{figure}

Note from \eqref{equiv_timing_error} that timing errors only occur when the input code changes, i.e., $\Delta x_n \neq 0$. In this case, the equivalent timing error is the average of the timing errors for each current cell that switches for the $x_{n-1} \rightarrow x_n$ code transition.\footnote{A derivation of the equivalent timing error, $T_\epsilon(n)$, is presented in \cite{bib:timing_errors}.} The error based on the equivalent timing error model is $e(t) = y(t) - y_\text{ideal}(t)$
and an example is illustrated in Fig. \ref{fig:error_pulses}. Note that $e(t)$ is comprised of narrow pulses with magnitudes $I_u |\Delta x_n|$ and durations $|T_\epsilon(n)|$. The magnitude of the net charge error introduced by the $x_{n-1} \rightarrow x_n$ transition is $Q_\epsilon(n) = I_u |\Delta x_n T_\epsilon(n)|$, i.e., the area under the error pulses. 
\subsection{Previous Analysis}
The average expected error power is computed in \cite{bib:timing_errors} as
\begin{subequations} \label{prev_analysis}
\begin{align} 
    \left<E[P_{e, \text{previous}}(n)] \right> &= \left<E\left[\left(\frac{Q_\epsilon(n)}{T_s}\right)^2\right] \right> \label{prev_analysis_1} \\ &= \left<E\left[\left(\frac{T_{\epsilon}(n)}{T_s}\right)^2 \Delta x_n^2 I_u^2\right] \right> \label{prev_analysis_2}\\ &= \frac{\sigma^2}{T_s^2} \, \left<|\Delta x_n| \right> \label{prev_analysis_3}
\end{align}
\end{subequations}
where $\left<\cdot \right>$ denotes discrete time averaging. We use \eqref{prev_analysis_3} to define the \textit{wideband} \ac{SDR} as
\begin{align} \label{wideband_sdr}
\text{SDR} = \frac{P_\text{sig}}{\frac{\sigma^2}{T_s^2} \left<|\Delta x_n|\right> }
\end{align}
since the denominator is comprised of the \textit{total} error power. Note that in \eqref{prev_analysis_1} it is implicitly assumed that the charge error occurs over the entire sample period, $T_s$. However, \textit{the charge error actually occurs over a small fraction of $T_s$}, as illustrated in Fig. \ref{fig:error_pulses}. In Section \ref{sec:analysis}, we carry out the analysis accordingly, resulting in a more accurate expression for the wideband \ac{SDR}. Lastly, the authors in \cite{bib:timing_errors} specialize \eqref{wideband_sdr} for low-frequency sinusoidal inputs with the error power limited to the first Nyquist band, resulting in an \ac{SDR} of
\begin{align} \label{sdr_nyq}
    \text{SDR}_{\text{Nyquist}} = \frac{A_1}{8f_1 f_s \sigma^2}
\end{align}
where $f_1 \ll f_s$ is the input frequency and $A_1 = \frac{1}{2} \, (2^M -1)$ is the input amplitude.\footnote{A more detailed derivation of \eqref{sdr_nyq} is presented in \cite{bib:doris_book}.}
\section{Analysis} \label{sec:analysis}
The error shown in Fig. \ref{fig:error_pulses} may be written as the sum of non-overlapping pulses $e_n(t)$, i.e., $e(t) = \sum_{n=-\infty}^\infty e_n(t)$ where 
\begin{align}\label{e_n(t)}
    e_n(t) = -\sgn(T_\epsilon(n)) \, I_u \Delta x_n \, \rect \left( \frac{t-(nT_s+T_\epsilon(n)/2)}{|T_\epsilon(n)|}\right)
\end{align}
and $\operatorname{rect}(t) = 1$ if $|t| \leq 1/2$ and $0$ if $|t|>1/2$.
Note that $e(t)$ is a random process, where the randomness comes from the equivalent timing errors, $T_\epsilon(n)$. The expected error power is
\label{appendix_A}
\begin{align}
    E[P_e] &= E\left[ \lim_{T\rightarrow \infty} \frac{1}{T} \int_{-T/2}^{T/2} \, e^2(t) dt \right] \nonumber \\ & = \lim_{N\rightarrow \infty} \frac{1}{(2N+1)} \nonumber\\ &~~~~~~ \times \sum_{n=-N}^N  E \left[\frac{1}{T_s} \int_{-(2N+1)T_s/2}^{(2N+1)T_s/2}  e_n^2(t) \, dt \right] \nonumber\\ &= \lim_{N\rightarrow \infty} \frac{1}{(2N+1)} \sum_{n=-N}^N E[A_n] \nonumber \\ &= \left< E[A_n] \right> \label{power_eqn}
\end{align}
where $A_n$ is a random variable defined by
\begin{align} \label{rv_im}
    A_n= \frac{1}{T_s} \int_{-(2N+1)T_s/2}^{(2N+1)T_s/2}  e_n^2(t) \, dt , \ \ \ |n| \leq N
\end{align}
\noindent Taking the expected value of \eqref{rv_im} and then averaging yields
\begin{align} 
    E[P_e]  &= \left<E \left[\frac{|T_{\epsilon}(n)|}{T_s} \Delta x_n^2 I_u^2 \right] \right>\label{E_I} 
\end{align}
Note from \eqref{equiv_timing_error} that $T_{\epsilon}(n) \sim N(0, \frac{\sigma^2}{|\Delta x_n|})$, $\Delta x_n \neq 0$. Therefore, $|T_{\epsilon}(n)|$ has a folded normal distribution \cite{bib:folded_normal} with mean $E[|T_{\epsilon}(n)|]  = \frac{\sigma}{|\Delta x_n|^{1/2}} \sqrt{\frac{2}{\pi}}$ which we substitute into \eqref{E_I}, yielding
\begin{align} 
    E[P_e]  = \frac{1}{T_s} \sqrt{\frac{2}{\pi}} \sigma I_u^2  \left< |\Delta x_n|^{3/2}  \right> \label{timing_error_power}
\end{align}
At this point, we can compare our analysis to that in \cite{bib:timing_errors}. Specifically, we observe that \eqref{E_I} and \eqref{prev_analysis_2} differ by a factor of $|T_\epsilon(n)|/T_s$ inside the expectation, i.e., the duty factor of the error pulses. In practice, $0 < |T_\epsilon(n)|/T_s \ll 1$, which means that this difference is nontrivial. The difference arises because the analysis in \cite{bib:timing_errors} implicitly assumes that the charge error is distributed over the entire sample period. In our analysis, we do not make this assumption. Using our analysis, the \ac{SDR} based on a sinusoidal input at full scale is
\begin{align}
    \text{SDR} &= 10 \log_{10} \left(\frac{P_\text{sig}}{E[P_e]}\right) \nonumber \\ &= 10 \log_{10} \left(\frac{(2^M-1)^2}{8 f_s \sqrt{\frac{2}{\pi}}  \left<|\Delta x_n|^{3/2} \right> }\right) - 10 \log_{10} \sigma \label{SDR_analysis}
\end{align}
where $P_\text{sig} = (\frac{I_u}{2} (2^M-1))^2/2$ and $E[P_e]$ comes from \eqref{timing_error_power}.  

\section{Simulations}\label{sec:simulatons}
\begin{figure}[h!]
\centering
\includegraphics[scale=0.32]{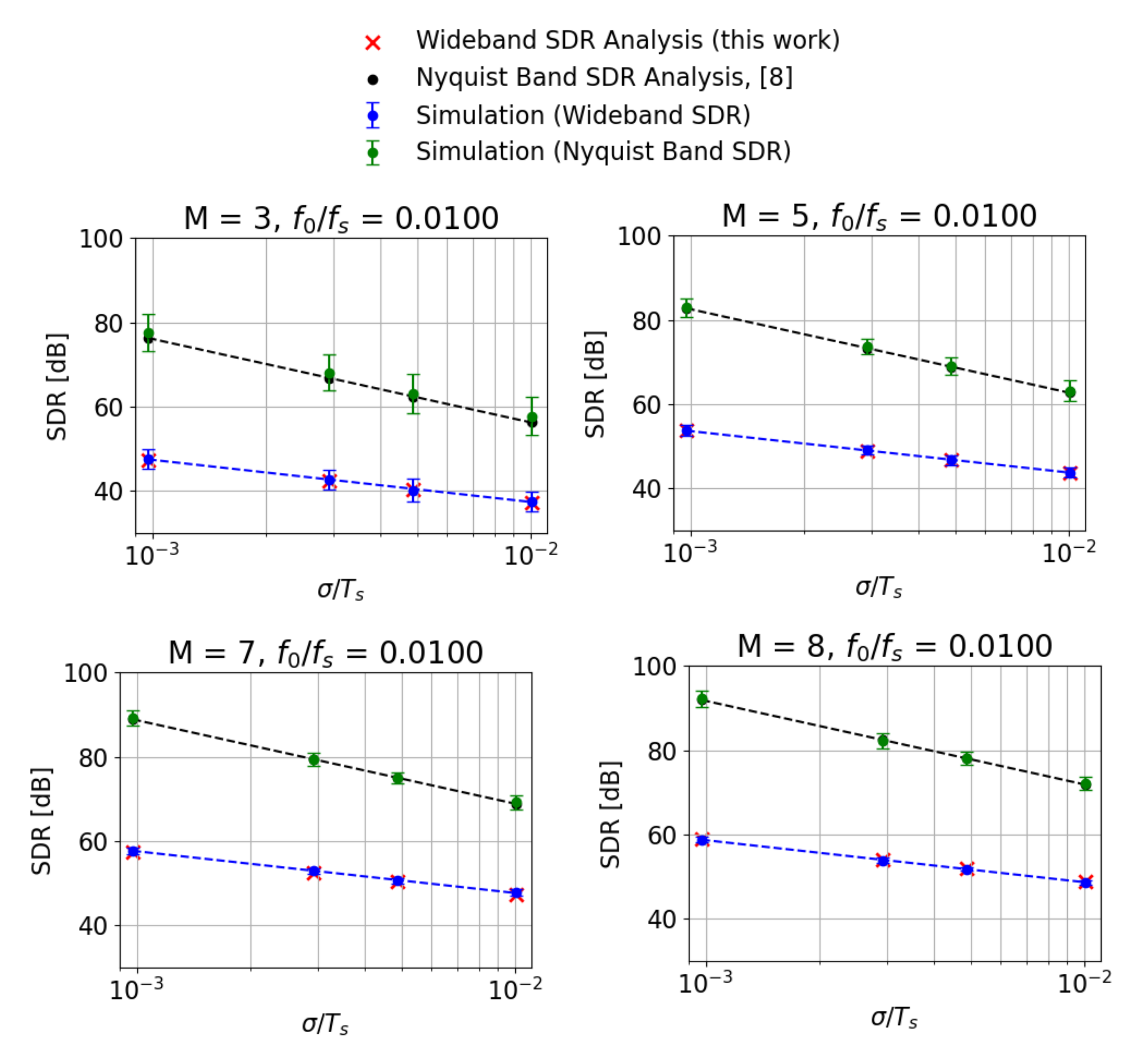}
\caption{SDR analysis vs. simulation for a single tone input, $f_0/f_s = 0.01$ -- the simulated data points are comprised of 50 independent runs of the behavioral model, where 95\% confidence intervals are shown.}
\label{fig:sdr_vs_res_v2}
\end{figure}

In this section, we simulate \ac{CS}-\ac{DAC}s using a behavioral model with an oversampling ratio (OSR) of 4096, i.e., one sample period, $T_s$, is represented by 4096 samples. The large OSR is so that we can capture timing errors that are a very small fraction of the sample period, e.g., we are interested in $\sigma/T_s$ down to $10^{-3}$. Note that using an $\text{OSR} = 4096$ is memory intensive, so we ran the simulations on a modern workstation with 128GB of RAM.
In Fig. \ref{fig:sdr_vs_res_v2}, we plot simulated and analytical results of the \ac{SDR} versus $\sigma/T_s$ for $M$-bit \ac{DAC}s ($M=3,5,7,8$). 
Note that the analysis in this work accurately captures the wideband \ac{SDR}, i.e., with no filter at the \ac{DAC} output. In contrast, the analysis in \cite{bib:timing_errors} accurately captures the Nyquist band \ac{SDR}, i.e., where there is brick wall filter at the  \ac{DAC} output with a cutoff frequency of $f_s/2$. Hence, the Nyquist band \ac{SDR} is substantially higher since it ignores the spectral content of the error beyond $f_s/2$. It is worth mentioning how the \ac{SDR} is extracted from the simulations. First, the behavioral model is run with zero timing errors to generate the ideal output, $y_\text{ideal}(n)$. Then, it is run with timing errors to generate the nonideal output, $y(n)$, resulting in an error of $e(n) = y(n) - y_\text{ideal}(n)$. For the wideband \ac{SDR}, the power of the sequence $e(n)$ is used. For the Nyquist band \ac{SDR}, the power spectral density of $e(n)$ is computed, and only the frequency components from DC to $f_s/2$ are included in the error power.
\begin{figure}[tb!]
\centering
\includegraphics[scale=0.38]{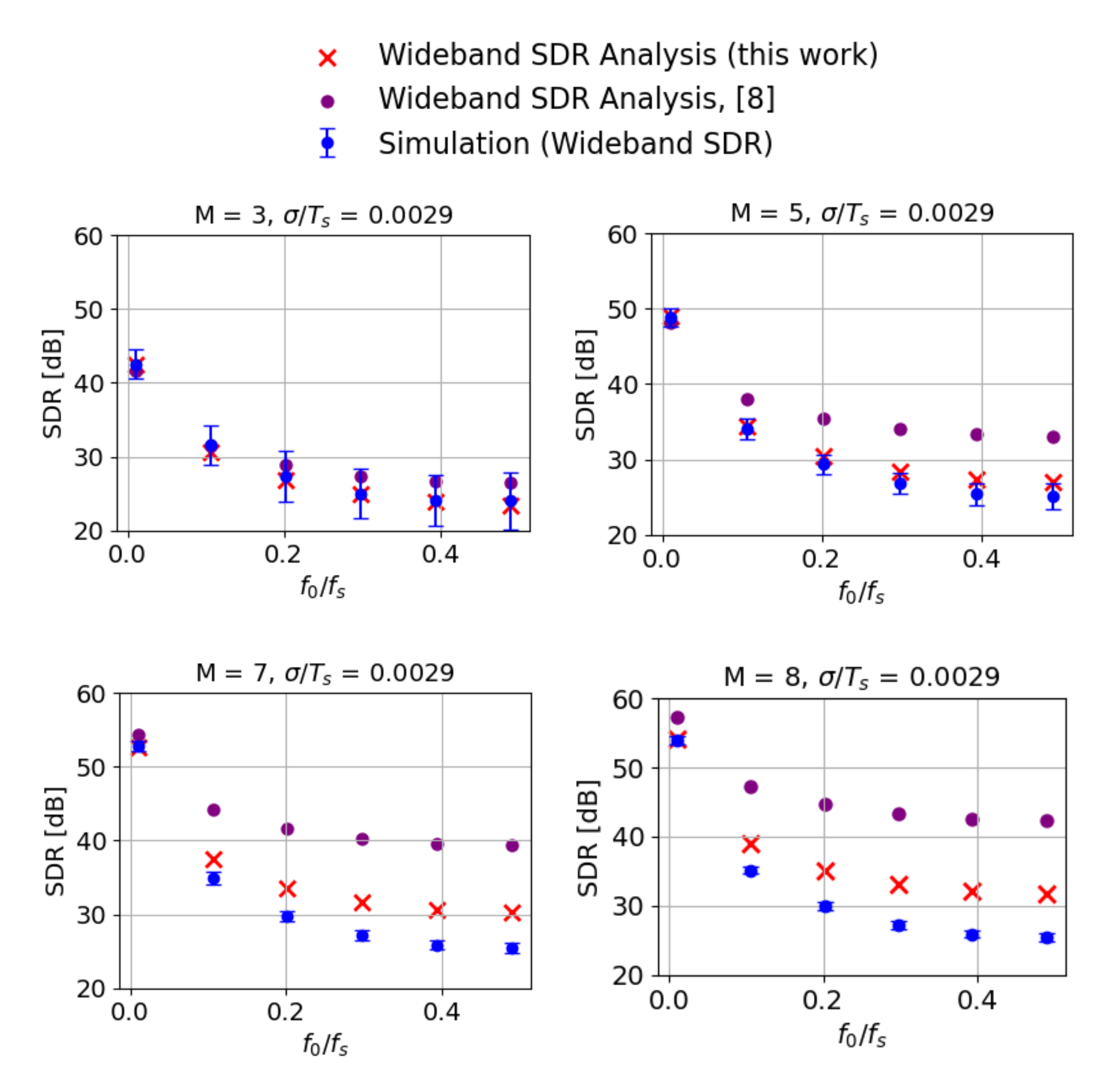}
\caption{\ac{SDR} analysis vs. simulation over frequency, $\sigma/T_s \approx \num{3e-3}$ -- the simulated data points are comprised of 50 independent runs of the behavioral model, where 95\% confidence intervals are shown.}
\label{fig:sdr_vs_freq}
\end{figure}
\begin{figure}[tb!]
\centering
\includegraphics[scale=0.4]{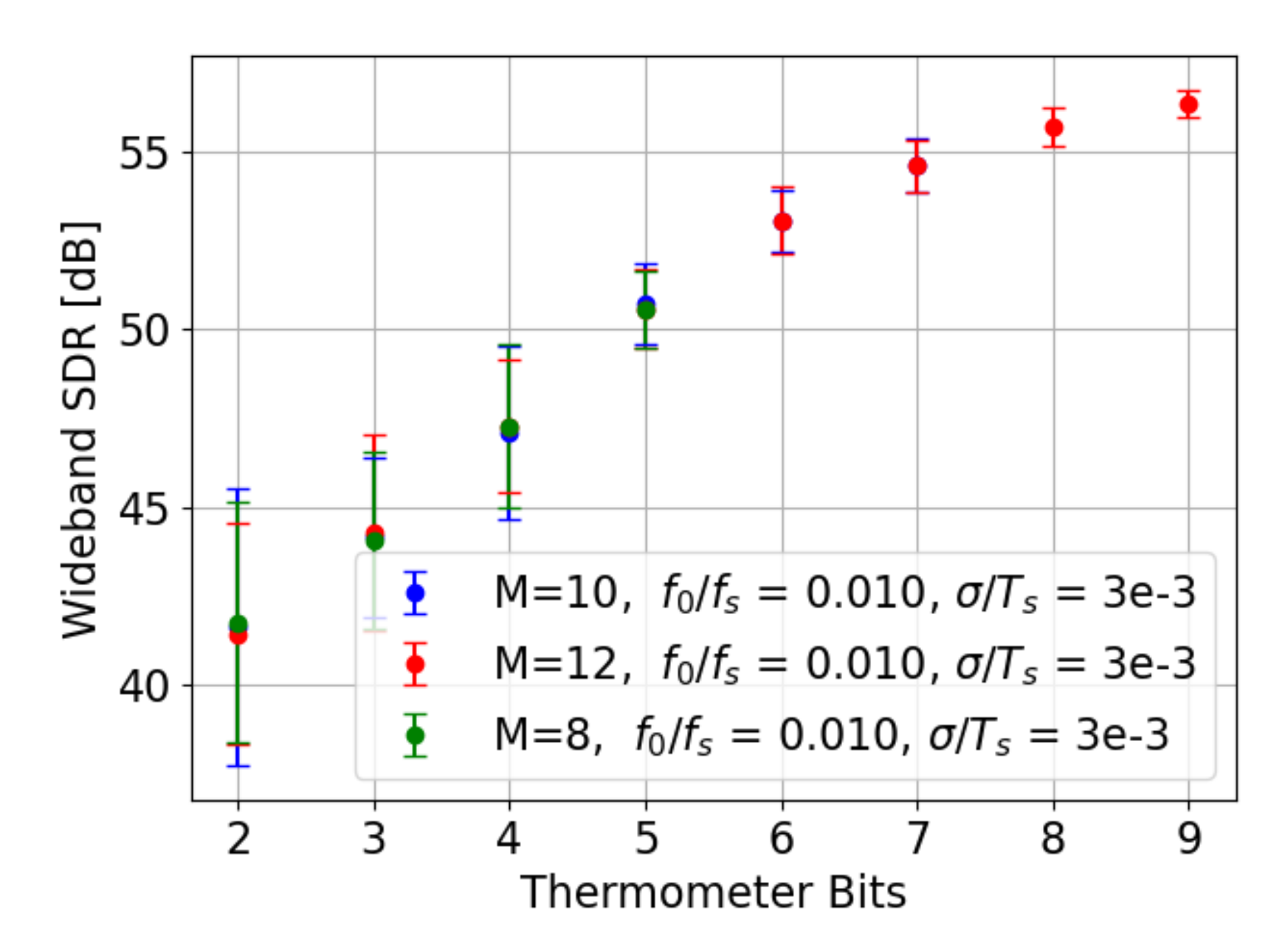}
\caption{Simulation of the wideband SDR versus thermometer bits for $M$-bit \ac{CS}-\ac{DAC}s ($M=8,10,12$) -- the simulated data points are comprised of 50 independent runs of the behavioral model, where 95\% confidence intervals are shown.}
\label{fig:sdr_vs_T}
\end{figure}

In Fig. \ref{fig:sdr_vs_freq}, we plot the wideband \ac{SDR} from simulation (in blue) and compare it with two different analyses over frequency. The red markers are from \eqref{SDR_analysis} (this work), and the purple markers are derived from the analysis in \cite{bib:timing_errors}, i.e., using \eqref{wideband_sdr} as the \ac{SDR}. Note that our analysis, in contrast to that in \cite{bib:timing_errors}, is in closer agreement with the simulations. This is because for practical values of $\sigma/T_s$, i.e., $0 < \sigma/T_s \ll 1$, the assumption in \eqref{prev_analysis_1} that the charge error is distributed over the entire sample period becomes less accurate. Since we do not make this assumption, there is a considerable difference between the two analyses for the small $\sigma/T_s$ considered in Fig. \ref{fig:sdr_vs_freq}. 

\begin{figure}[tb!]
\centering
\includegraphics[scale=1.1]{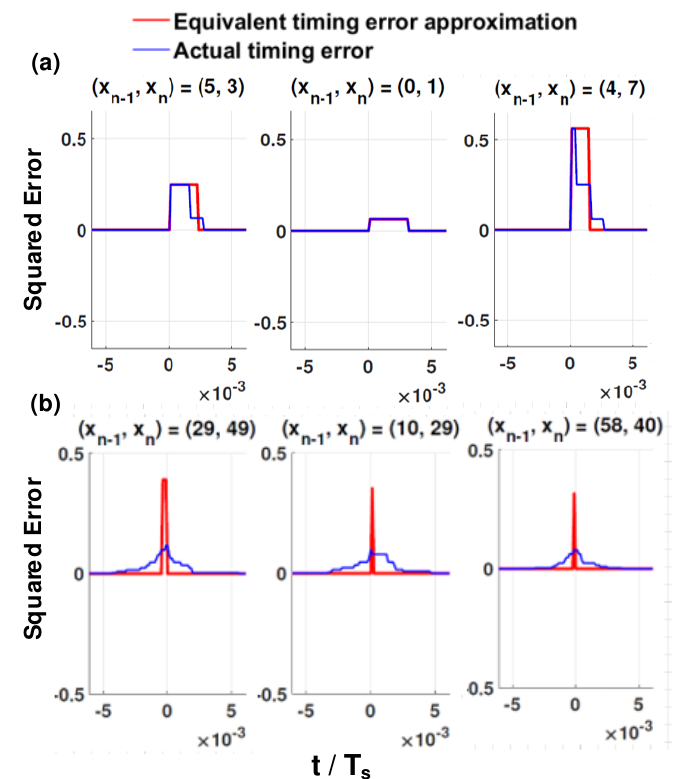}
\caption{Squared error for various code transitions $x_{n-1} \rightarrow x_n$ in an $M$-bit \ac{CS}-\ac{DAC}, $f_0/f_s \approx 0.11$, $\sigma/T_s \approx \num{3e-3}$. For each case, the nominal switching instant $nT_s$ is aligned with $t=0$, (a) $M=3$ and (b) $M=6$. }
\label{fig:timing_errors_qualitative}
\end{figure}

Referring again to Fig. \ref{fig:sdr_vs_freq}, note that our analysis diverges from the simulation as the number of bits, $M$, is increased (by up to 6.3dB for $M=8$ and $f_0/f_s \approx 0.5$). This divergence is caused by the breakdown of the equivalent timing error model for larger values of $M$ at high frequencies. This is qualitatively shown in Fig. \ref{fig:timing_errors_qualitative}(a) and Fig. \ref{fig:timing_errors_qualitative}(b), where we illustrate the squared error for various code transitions for 3-bit and 6-bit \ac{DAC}s, respectively.
Note that the squared errors in the 3-bit cases closely resemble rectangular pulses, i.e., they are well-suited for approximation via equivalent timing errors. In contrast, the squared errors for the 6-bit cases change gradually in the vicinity of the switching instant. 

We also used the behavioral model to investigate partially-segmented \ac{DAC}s. We considered $M$-bit \ac{DAC}s with the first $T$ bits (MSBs) thermometer-decoded into $2^T-1$ unit elements and binary weights for the remaining $M-T$ bits (LSBs). In Fig. \ref{fig:sdr_vs_T}, we plot simulations of the wideband \ac{SDR} versus $T$ for $M$-bit partially-segmented \ac{DAC}s ($M=8,10,12)$. For $T<6$, the \ac{SDR} increases by approximately 3dB/bit. However, the \ac{SDR} eventually saturates, i.e., the improvement gets smaller for each bit added when $T\geq6$. For example, going from $T=8$ to $T=9$ yields only a 1dB increase in \ac{SDR}. Lastly, it should be noted that if $M$ is sufficiently large, e.g., $M\geq8$, then increasing it further does not improve the \ac{SDR}, i.e., the impact of quantization on the timing errors becomes negligible. 
\section{Conclusion}\label{sec:conclusion}
In this work, we presented analysis of the wideband \ac{SDR} due to timing errors in fully-segmented \ac{CS}-\ac{DAC}s, which was validated using a behavioral model and proven to be significantly more accurate than the previous analysis. In addition, we used the model to characterize the \ac{SDR} for partially-segmented architectures. Thus, this work provides a method for accurately specifying error tolerance in the circuit design to achieve a given \ac{SDR}. A useful extension would be to improve the model accuracy for high-resolution \ac{DAC}s at high-frequency operation. This may be done by substituting the equivalent timing error model with one that more accurately captures the error pulse characteristics.



\bibliographystyle{IEEEtran}
\bibliography{bibliography}
\appendices


\end{document}